\documentclass[10pt,journal,compsoc]{IEEEtran}

	\usepackage[utf8]{inputenc} 
	\usepackage[iso]{datetime}
	\usepackage{amssymb}
	\usepackage{enumerate}
	
	\usepackage[a4paper]{geometry}
	
	\usepackage{xcolor}
		\definecolor{darkred}{rgb}{0.8,0.1,0.1}
		\definecolor{darkgreen}{rgb}{0,0.5,0}
		\definecolor{darkblue}{rgb}{0,0,0.5}
		\colorlet{RED}{red}
	
	\usepackage[colorlinks=true,linkcolor=red,urlcolor=blue,citecolor=red]%
		{hyperref}
	\usepackage{graphicx,epstopdf}
	\graphicspath{{../common/figures/}}
	\usepackage{subcaption}

	\newcommand{\reffig}[1]{Fig.~\ref{#1}}
	\newcommand{\refeq}[1]{Eq.~\ref{#1}}
	
	\newcommand{\refsec}[1]{Sec.~\ref{#1}}

	\usepackage{amsmath, amssymb,amsfonts,amsthm}
	\theoremstyle{plain}

	\theoremstyle{definition}

	\theoremstyle{remark}

	\newcommand{\hAbs}[1]{\ensuremath{\left \lvert \, #1 \, \right \rvert} } 
	
	\newcommand{\hArgmin}[2]{\underset{#1}{\operatorname{arg \, min}}\;#2}
	\newcommand{\hArgmax}[2]{\underset{#1}{\operatorname{arg \, max}}\;#2}

	\usepackage{changes}
	
	\definechangesauthor[name=Haluk Bingol,color=purple]{hb}
	\definechangesauthor[name=Meliksah Turker,color=blue]{mt}
	
\begin{document}
\title{
	Forgiveness is an Adaptation\\
	in Iterated Prisoner’s Dilemma with Memory
}

\author{
	Meliksah~Turker,
	Haluk~O.~Bingol
	\IEEEcompsocitemizethanks{
		\IEEEcompsocthanksitem M.~Turker and H.~O.~Bingol  are with Bogazici University.
	}%
}


\IEEEtitleabstractindextext{%
	\begin{abstract}
		The Prisoner's Dilemma is used to represent many real life phenomena
		whether from the civilized world of humans or from the wild life of the other living.
		Researchers working on iterated prisoner's dilemma (IPD) with limited memory
		inspected the outcome of different forgetting strategies in homogeneous environment,
		within which all agents adopt the same forgetting strategy at a time. 
		In this work, with the intention to represent real life more realistically,
		we improve existing forgetting strategies, offer new ones, and conduct experiments
		in heterogeneous environment that contains mixed agents and compare the results
		with previous research as well as homogeneous environment.
		Our findings show that the outcome depends on the type of the environment,
		and is just the opposite for homogeneous and heterogeneous ones, opposing the existing literature in IPD. 
		Consequently, forgetting and forgiving defectors is the supreme memory management strategy 
		in a competitive, heterogeneous environment.
		Therefore, forgiveness is an adaptation.
	\end{abstract}
	
}

\maketitle


\section{Introduction}

Memory is a valuable, limited resource in life.
It is the source that is used to store the data 
that feeds the decision making process and for that reason, 
it is crucial to keep valuable, informative ones
while throwing out unrelated, non-value creating ones since it is limited by nature. 
But what are the good, valuable, informative data to choose and keep among many?~\cite{%
	bingol2008fame}.

\subsection{Prisoner's Dilemma}

Prisoner's dilemma is used to represent a range of real life phenomena such as 
economics~\cite{%
	kreps1990game}, 
commerce~\cite{%
	cetin2014CMP},
nature and wildlife~\cite{%
	nowak2006evolutionary}. 
It represents cooperation and competition among agents. 

In prisoner's dilemma, agents come together to play and each agent 
either cooperates or defects without knowing opponent's action~\cite{%
	axelrod1981evolution}.
This can result in four different outcomes. 
If they both cooperate, they both receive the reward payoff $R$. 
If one cooperates while the other one defects, 
the cooperating agent receives the sucker payoff $S$ and 
defecting one receives temptation payoff $T$. 
If they both defect, then each agent receives punishment payoff $P$. 
For such a game to be prisoner's dilemma, it has to satisfy two conditions:
(i)~$S < P < R < T$, and 
(ii)~$S + T < 2R$.

Natural result of this schema is, 
defecting is always the rational decision, regardless of opponent's action. 

However, this so called \emph{rationality} changes in
iterated prisoner's dilemma with memory, where
\begin{enumerate}[i.]
	\item 
	prisoner's dilemma game is played multiple times between agents~\cite{%
		axelrod1981evolution},
	
	\item 
	agents remember the past actions of their opponents~\cite{%
		axelrod1987evolution},

	\item 
	agents refuse to play against defectors~\cite{%
		tesfatsion1997refusal}.
\end{enumerate}

As long as there exists free space in their memories,
agents can keep track of all opponents.
If there is not enough memory capacity,
some memory management strategy is needed~\cite{%
		cetin2014CMP}.
Consider an iterated prisoner's game with $N$ agents.
Let $M$ be the memory capacity of an agent.
If $M < N$, 
Agent ``remembers'' the first $M$ distinct opponents 
but there is no space left for the next opponent.
Once an agent's memory is full, 
it opens space by forgetting a known opponent
according to its ``forgetting strategy''.

In this work, we review existing memory management strategies
in IPD literature, improve them and offer novel ones and report
the outcome for two different type of environments,
namely \emph{homogeneous} and \emph{heterogeneous}.

\section{Related Work}

Prisoner's Dilemma (PD) has developed a wide appeal towards 
the end of the 20th century.
Even though defection seems to be the rational choice regardless of the
opponent's action, nature has found ways to promote cooperation
over defection.
In order to resolve this seemingly contradictory phenomenon,
researches have studied PD to understand it
and find the circumstances that make cooperation more favorable
than defection.

In 1980, Axelrod organized a computer simulation based tournament 
where agents of various strategies would play against agents of other strategies 
and themselves.
Among the various strategies such as always defect, random, always cooperate,
go-by-majority and Tit-for-Tat, the winner was Tit-for-Tat, 
that always cooperates in the first round,
then copies the opponent's action from the previous round in the other rounds.
Axelrod analyzed and published the results in his 1984 book~\cite{%
	robert1984evolution}.
He observed that the success in evolutionary games requires being
(i)~nice,
(ii)~provocable,
(iii)~not envious,
and
(iv)~not too clever.

Axelrod further investigated the ways to promote cooperation
by introducing iteration and memory\cite{%
	axelrod1981evolution,
	axelrod1987evolution}.

Even though "being nice" is observed to be beneficial
against cooperators, it requires a mechanism to be protected
against defectors.
Ref~\cite{tesfatsion1997refusal} proposed ``refusal'' in game playing, 
where an agent can refuse to play against a known defector,
thus protecting itself.

Ref~\cite{cetin2014CMP} investigated memory management strategies
in IPD with limited memory,
where agents are unable to keep track of every opponent in the environment,
therefore need a memory management strategy.
It is reported that
in an environment consisting of pure cooperator and pure defector agents,
where all agents use the same forgetting strategy,
the group of agents that perform the best are the ones
that adopt ``forget only cooperators'' strategy.

In a follow up work~\cite{%
	cetin2016dose},
pure cooperators and defectors are relaxed by 
probabilistic cooperators and defectors, where each agent has an internal parameter, 
called \emph{cooperation probability} $\rho$,
and cooperates according to this probability.

With the introduction of cooperation probability, 
agents are no longer radicals that always cooperate or defect,
but behave somewhat stochastically.
This change brought the idea of \emph{perceived cooperation ratio},
that is how an agent knows another based on its past actions against 
the focal agent.

Following this path, ref~\cite{cinar2020getting} proposed a model
where agents ask for recommendations from known agents
instead of simply playing the first round
when they are matched against unknown opponents.

Ref~\cite{ma2021limited} investigated the effect of memory size
on cooperation and attempted to set an ideal memory size,
in a different setup, 
where agents of three main strategies existed.
They are always cooperate, always defect and lastly,
cooperate against cooperator and defect otherwise.

In another branch in IPD literature, Press and Dyson~\cite{%
	press2012iterated}
discovered a new group of strategies called Zero-determinant (ZD),
which forces the average payoff between agents to be linearly related,
effectively setting an average payoff for the opponent
regardless of the opponent's strategy.

Another group of researchers~\cite{%
	glynatsi2020using,%
	hilbe2017memory%
}
investigated memory-$n$ strategies in IPD,
where an agent's decision to cooperate or defect is affected by
the opponent's actions against the focal agent in the last $n$ rounds.

Ref~\cite{%
	ueda2021memory}
extended the ZD strategies and applied them to 
memory-two and memory-$n$ strategies in repeated games.

Ref~\cite{dal2019strategy} inspected the successful strategies
in infinitely repeated prisoner's dilemma and classified them under 
three fundamental groups,
(i)~always defect,
(ii)~Tit-for-Tat,
(iii)~grim.

\section{Method}

Here, we extend IPD with limited memory by
(i)~improving forgetting strategies,
(ii)~offering new forgetting strategies,
(iii)~working on heterogeneous environment,
and
(iv)~using \refeq{eq:t_ij_new} for perception.

\textbf{Memory.}
Agents have a memory of their past experiences.
An agent $i$ remembers who it played with 
and for how many times its opponent $j$ cooperated and defected against it,
$c_{ji}, d_{ji}$, 
respectively.
In any given time, an agent can remember up to $M$ different opponents
regardless of how many games it played against those opponents.
Agents are homogeneous in terms of memory size $M$,
thus all agents in the environment has the same memory capacity.
\emph{Memory ratio} is the ratio of agents that an agent can keep in its memory,
$\mu = M / N$.
Once the memory of an agent is full and it will play a game against 
an unknown opponent, it chooses a known opponent to forget
according to its forgetting strategy, 
discussed in \refsec{sec:Forgetting Strategies}, 
and removes it.

\textbf{Cooperation Probability.}
As life is not black and white, but a continuous gradient of gray,
so are our agents' cooperation probability $\rho$.
We consider 21 cooperation probabilities 
starting at pure defector $\rho = 0$, 
ending at pure cooperator $\rho = 1$,
incremented by $0.05$, given by 
\begin{align} 
	\rho_{k} = 0.05 k, \quad k = 0, 1, \dotsc, 20.
	\label{eq:gradient_coop_prob}
\end{align}

This way, 
we improve the pure cooperator, 
i.e., $\rho = 0$, 
and 
pure defector, 
i.e. $\rho = 1$, 
agents of 
ref~\cite{cetin2014CMP}
by using a fixed gradient of cooperation probabilities instead of
setting $\rho$ randomly for each agent~\cite{%
	cetin2016dose}.

\textbf{Perceived Cooperation Ratio.}
Agents have an opinion about other opponents according 
to their past experience,
namely the perceived cooperation ratio.
In order to perceive, an agent $i$ keeps two numbers in its memory,
the number of times an opponent $j$ cooperated and defected,
$c_{ji}$ and $d_{ji}$, 
respectively.
The perceived cooperation ratio is defined as
\begin{align}  
	t_{ij} = \frac{c_{ji} + 1}{(c_{ji} + 1) + (d_{ji} + 1)}
	\label{eq:t_ij_new}
\end{align} 
which effectively gives greater weights to larger samples~\cite{%
	cetin2016dose}.

An agent is perceived as \emph{cooperator} if 
$t_{ij} > 0.5$;
\emph{defector} otherwise,
where $t_{ij}$ is evaluated as in \refeq{eq:t_ij_new}.
This is the cooperator-defector definition used throughout this work.

\textbf{Game Playing.}
Agents come together to play prisoner's dilemma game.
In each round, two agents are selected uniformly at random. 
Each agent decides whether it will play or not, depending on the opponent. 
If an agents ``knows'' its opponent as defector, 
it refuses to play and the round results with no game played. 
If it knows the opponent as cooperator, or does not know the opponent at all, it plays. 
A game is played only if both agents agree to play. 

At the end of the game both players update their memories.
If the opponent $j$ is already in the memory,
either $c_{ji}$ or $d_{ji}$ is incremented
according to the opponent's action. 
If $j$ is not in the memory, an empty space is used for $j$.
If there is no empty space,
then agent $i$ uses its forgetting strategy to free a space for $j$.

\subsection{Forgetting Strategies}
\label{sec:Forgetting Strategies}

We introduce five new memory strategies in addition to forget randomly(FR)~\cite{%
	cetin2014CMP}.

\textbf{Forget randomly (FR).}
In this strategy, the agent chooses an opponent from its memory 
uniformly at random and forgets it.
This is the strategy used in the previous works.

\textbf{Forget most cooperator first (FMC).}
In FMC strategy, the agent chooses the opponent with the highest
perceived cooperation ratio from its memory and forgets it.
That is,
\[
	\hArgmax
		{j \in \mathcal{M}_{i}}
		{ \{t_{ij}\} }
\]
where $\mathcal{M}_{i}$ is the memory content of agent $i$.

\textbf{Forget most defector first (FMD).}
Conversely, in FMD, the agent forgets 
the opponent with the lowest perceived cooperation ratio.

Note that these two strategies are improved versions of 
``forget only cooperators'' and ``forget only defectors'' strategies of ref~\cite{%
	cetin2014CMP},
where agent to be forgotten was chosen randomly among its group
with no regards to how strong a cooperator or a defector it was.
Considering that the agents in this work are not pure cooperators or pure defectors,
we believe these versions are more realistic and better than 
randomly selecting a cooperator or defector.

\textbf{Forget most unpredictable first (FMU).}
In FMU, the agent forgets the shadiest opponent first,
that is, 
whose perceived cooperation ratio is closest to 0.5 given by
\[
	\hArgmin
		{j \in \mathcal{M}_{i}}
		{ \{ \hAbs{t_{ij} - 0.5} \} }.
\]
Justification of this strategy is that 
remembering these agents 
should not bring any useful information and
must be a waste of memory 
since their actions are similar to a coin toss.

\textbf{Forget least played first (FLP).}
In FLP,
the agent forgets the opponent 
with the least games it played against.
That is,
\[
	\hArgmin
		{j \in \mathcal{M}_{i}}
		{ \{ c_{ij} + d_{ij} \} }.
\]
The idea is to keep agents with larger samples 
because it would bring statistical consistency 
while forgetting agents with small samples.

\textbf{Forget most played first (FMP).}
Conversely,
FMP implies forgetting the opponent with the largest sample first.
Thus, the definition of minority and majority are different than ref~\cite{cetin2014CMP}.

\textbf{Evaluation Metric.}
In evolutionary dynamics~\cite{%
	nowak2006evolutionary}, 
the payoff represents \emph{relativity},
that is, \emph{compared to} other agents in the environment.
Hence, 
their relative success can be measured with their payoff ratio defined as follows.
Let $\mathcal{A}$ be the set of all agents.
Let $\mathcal{B}  \subseteq \mathcal{A}$ be the set of agents that we are interested in.
\emph{Payoff ratio} of agents in  $\mathcal{B}$ is given as
\[
	\phi_{\mathcal{B}} 
	= \frac
		{\overline{P_{\mathcal{B}}}}
		{\overline{P_{\mathcal{A}}}},
\]
where $\overline{P_{\mathcal{B}}}$ is
\[
	\overline{P_{\mathcal{B}}} 
	= \frac{1}{\hAbs{\mathcal{B}}} 
	   \sum_{i \in \mathcal{B}}{\textnormal{payoff}(i)}.
\]
Note that we have modified the payoff ratio defined in ref~\cite{cinar2020getting}.
This modification does not change the result for homogeneous environment
but allows comparison of a group to all others,
including cooperators that use other forgetting strategies,
in case of heterogeneous environment.

It should also be noted that the agents in $\mathcal{B}$ are doing better than
the average population when $\phi_{\mathcal{B}} > 1$.

\section{Experiments}

We begin conducting experiments by creating and placing agents into an environment.
An environment consists of $N = \hAbs{\mathcal{A}} = 126$ agents 
with a given memory ratio $\mu$ that is constant for all agents in it.
We run the system for every value of
$ \mu \in \{0.05m \mid m = 0, 1, \dotsc, 20 \} $

The choice of $N = 126$ comes from $21 \cdot 6$,
where 
21 is the number of discrete $\rho_{k}$ values
according to \refeq{eq:gradient_coop_prob},
and 6 is the number of forgetting strategies inspected in this work.
In a single experiment, agents play a total of ${N \choose 2} \tau$ games.
We set $\tau = 30$.
Thus every agent plays with each other 30 times on average~\cite{%
	cetin2014CMP},
which we call a \emph{realization}.
We report the average of 50 realizations.

\begin{figure*}[!htp]
	\begin{subfigure}[b]{\linewidth}
		\centering
		\includegraphics[width=\linewidth]%
			{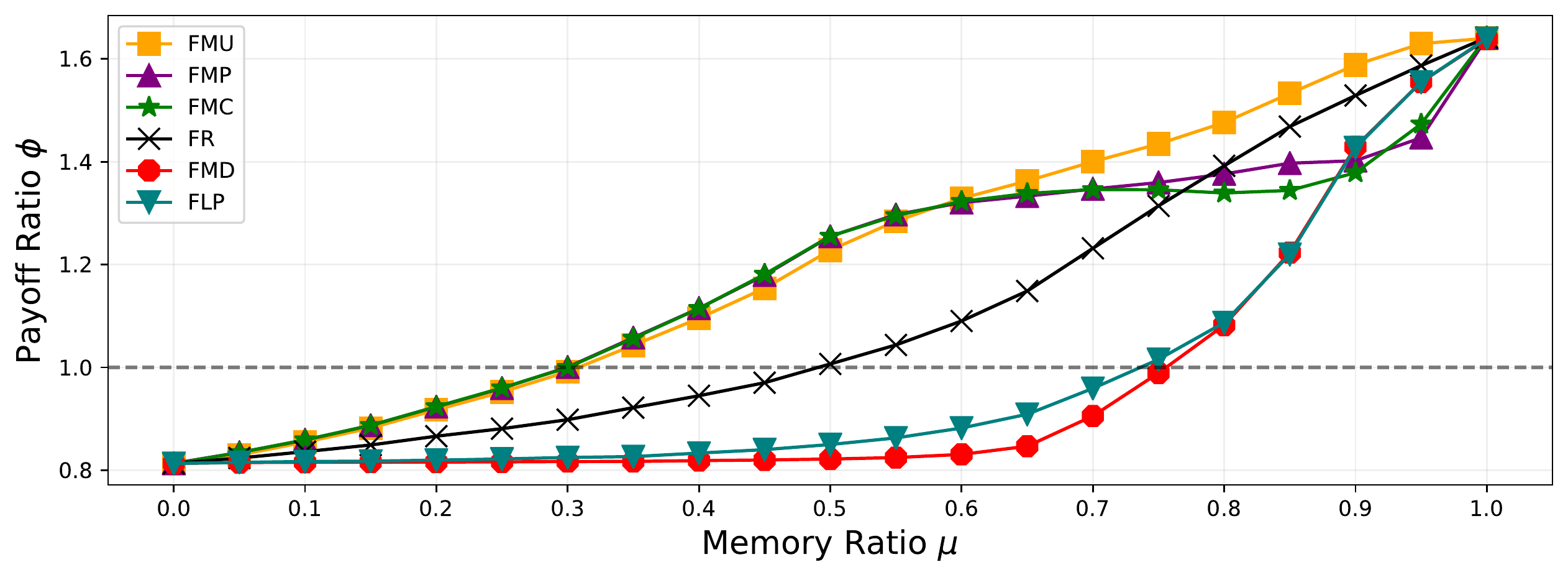}
		\caption{Homogeneous}
		\label{fig:homogeneous}
	\end{subfigure}%
	\\
	\begin{subfigure}[b]{\linewidth}
		\centering
		\includegraphics[width=\linewidth]%
			{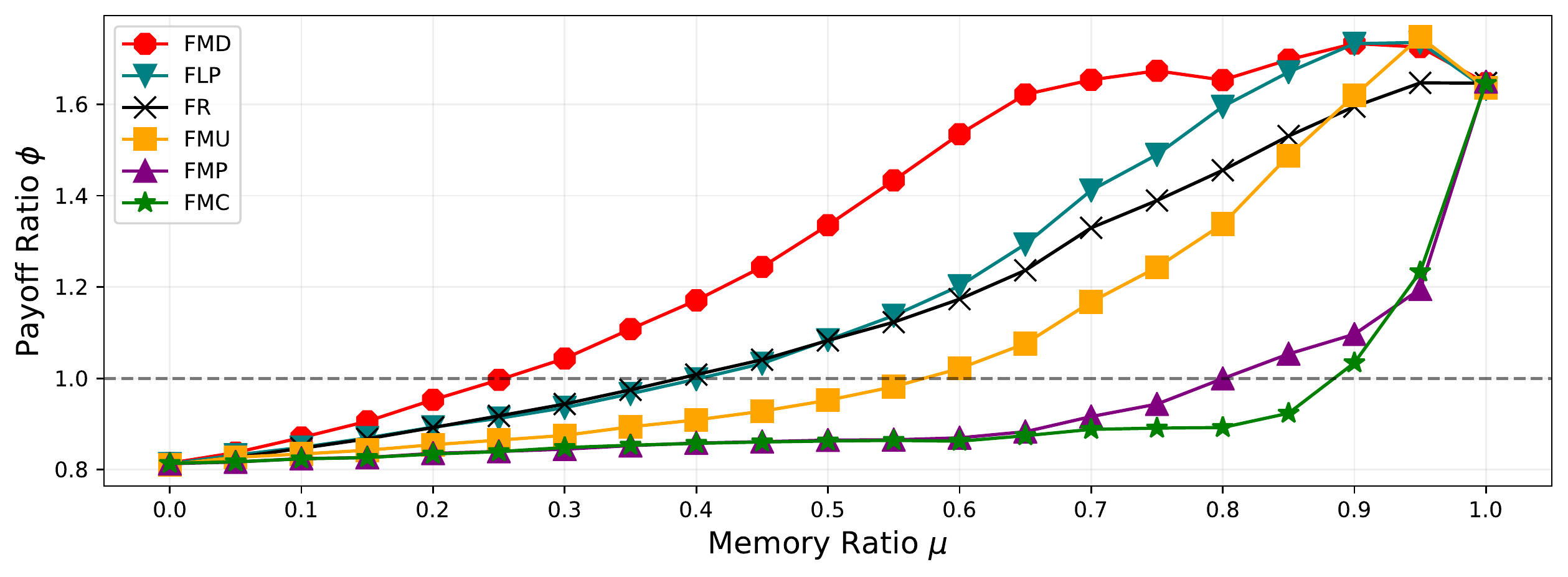}
		\caption{Heterogeneous}
		\label{fig:heterogeneous}
	\end{subfigure}
	\caption{
		Payoff ratio $\phi$ of cooperators, that is, 
		agents with $\rho > 0.5$ in homogeneous and heterogeneous environments.
	} 
	\label{fig:homohetero}
\end{figure*}

\subsection{Homogeneous Environment}

\emph{Homogeneous environment} consists of agents with single forgetting strategy $s$.
This is the setting used in refs~\cite{%
	cetin2014CMP,%
	cinar2020getting}.
With 126 agents, 21 distinct $\rho$ values and single $s$,
there are 6 agents of each $\rho$ value in homogeneous environment.

In this schema, agents of a single $s$ and a constant $\mu$ value
play against each other and payoff ratio $\phi$ of cooperators is reported 
as a single point in \reffig{fig:homogeneous}.
This makes a single experiment and every point in the \reffig{fig:homogeneous}
another, independent experiment.
Conducting experiments for every distinct $\mu$ value and six $s$,
we obtain the entire figure.

\subsection{Heterogeneous Environment}

Even though homogeneous environment allows controlled experiments
via direct comparison of cooperators and defectors that use the same forgetting strategy,
it does not represent a more competitive, realistic environment.
Therefore we introduce \emph{heterogeneous environment},
in which agents of all strategies coexist and
compete with each other via game playing.

The natural outcome of this schema is that
there is a single agent for each $\rho$ and $s$.

Since the environment now consists of agents with different forgetting strategies,
single experiment results in six payoff ratio $\phi$ values, 
one for cooperators of each $s$.
Consequently, we report the results of a single experiment as six data points in \reffig{fig:heterogeneous}, 
where they share a single $\mu$ value.
Conducting this for all $\mu$ values, we obtain the entire figure.

\begin{figure*}[!htp]
	\centering 
	\includegraphics[width=15cm]%
		{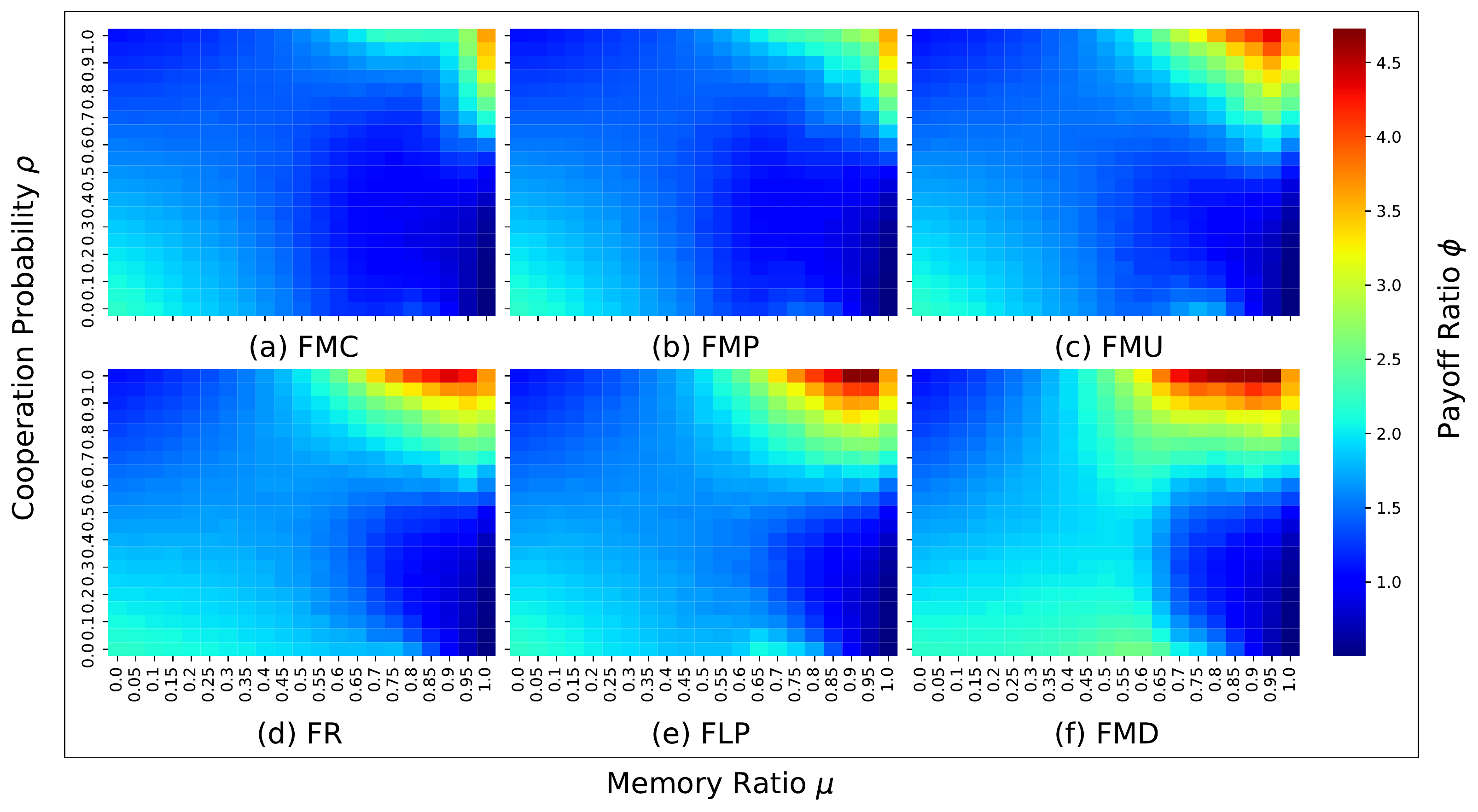}
	\caption{Individuals payoffs in heterogeneous environment} 
	\label{fig:heatmap}
\end{figure*}

\subsection{Individual Payoffs}

Lastly, we inspect the payoffs of individual agents in heterogeneous 
environment in \reffig{fig:heatmap}.
There are six sub-figures, each for the corresponding forgetting strategy,
and $21 \cdot 21 = 441$ data points, 
that arise as result of 21 distinct $\mu$ and $\rho$ values.
Each data point shows the payoff of an agent with corresponding 
$\mu$ and $\rho$.
This makes it possible to inspect the payoff ratio $\phi$ for both cooperators
and defectors in high resolution.

\section{Results and Discussion}

\subsection{Homogeneous vs. Heterogeneous}

In \reffig{fig:homogeneous} it is observed that:
(i)~For $\mu < 0.75$, 
the order of superiority (FMC $>$ FR $>$ FMD) 
is similar to the results (FC $>$ FR $>$ FD) of earlier works~\cite{%
	cetin2014CMP,%
	cinar2020getting}.
(ii)~This changes after $\mu =0.75$ and $\mu = 0.90$,
where FR and then FMD surpasses FMC, respectively.
(iii)~FMU performs the best in terms of consistency,
considering the dramatic changes that occur for others at higher $\mu$ values.
(iv)~Interestingly, 
FMC and FMP move very closely for $\mu < 0.7$.
So do FMD and FLP.

Compared to \reffig{fig:homogeneous},
drastic changes are observed in \reffig{fig:heterogeneous}:
(i)~The best performing forgetting strategy is consistently FMD,
which was the worst in the homogeneous case.
It is followed by FLP.
(ii)~FMC,
which was a superior strategy in \reffig{fig:homogeneous} and
refs~\cite{cetin2014CMP,cinar2020getting},
is now the worst performing strategy.
(iii)~FMU, which was the top strategy in \reffig{fig:homogeneous},
is now a mediocre strategy.

The results in \reffig{fig:heterogeneous} change completely 
compared to \reffig{fig:homogeneous}.
Previously top performing forgetting strategies in refs~\cite{%
	cetin2014CMP,%
	cinar2020getting}
and homogeneous environment have become the worst performing ones
in heterogeneous environment.
The agents that forget defectors perform the best by far.

The reason behind this drastic change is the opponents agents compete against.
A group of agents that adapt a weak strategy can obtain good payoff ratio
in an environment full of weak/weaker agents.
However when placed into an environment with agents with stronger strategy,
game changes and they obtain a poor payoff ratio.
This is the relativity in evolutionary dynamics.

Consequently, homogeneous environment only reflects
the success of cooperators that use a forgetting strategy 
against defectors of the same forgetting strategy,
whereas heterogeneous environment reveals
more comparable results.

We conducted the same set of experiments
(i)~using FC and FD instead of FMC and FMD respectively,
(ii)~setting cooperation probabilities $\rho$ according to ref~\cite{%
	cinar2020getting}.
It is observed that the results,
which are not reported as figures due to lack of space, 
are quite similar to \reffig{fig:homohetero}.
Therefore our findings are not result of these alterations,
but because of heterogeneity.

\subsection{Individual Payoffs}

Inspection of individual level payoff ratios in \reffig{fig:heatmap} reveals more details.

(i)~The payoffs for the columns of $\mu = 0$ and $\mu = 1$ are the same.
In the case of $\mu = 0$, an agent has no memory so it cannot forget
what it does not know.
In the case of $\mu = 1$, an agent is able to keep track of everyone in the environment
so it does not need to open space by forgetting.
Consequently the first and last column are identical for all forgetting strategies
because forgetting does not occur in these cases.

(ii)~The upper right side of each sub-figure is the warmest, that is,
cooperators outperform defectors in case of significant memory.
Conversely, defectors outperform cooperators in case of low memory,
at the lower-left of each sub-figure.
However, the difference is not as significant as 
cooperators outperforming defectors at the upper right corners.
Interestingly, for strategies FMU, FR, FLP and FMD,
the warmest region is not where $\mu = 1$ but before that.
The reason is forgetting does not occur for $\mu = 1$ and 
this is elaborated in section \ref{subsec:ForgivenessIsAnAdaptation}.

(iii)~Supremacy of FMD is observed with largest and warmest regions in
high $\mu$ and $\rho$.
Moreover, defectors of FMD also outperform defectors of other forgetting strategies,
as seen in a large turquoise region of low $\mu$ values.
This shows that FMD is the ultimate forgetting strategy not only for cooperators,
but also for defectors.

(iv)~A strange phenomenon is observed in FMU and FLP sub-figures.
As $\mu$ increases from 0 to higher values, 
payoff ratio of defectors first decrease, 
then increase again around $\mu = 0.7$.
It appears that higher $\mu$ benefits defectors of these two forgetting strategies
against defectors of other forgetting strategies.
The case is also similar for FMD, with
higher $\mu$ benefiting defectors.

\subsection{Forgetting similar agents}

Pairs of (FMC, FMP) and (FMD, FLP) move very closely in \reffig{fig:homogeneous}.
This is also the case in \reffig{fig:heterogeneous},
where FMC and FMP move closely at the bottom and
FMD and FLP move less closely at the top of the figure.
This is because these two pairs of forgetting strategies end up forgetting very similar agents.
Since an agent refuses to play with an opponent once it knows it as a defector,
it is always the defectors that are played against the least.
Conversely, an agent will continue to play with an opponent it knows as cooperator,
and this will increase the number of games played against that opponent.
Consequently, agents that use the forgetting strategies FMC and FMP tend to forget cooperators,
and agents that use the forgetting strategies FMD and FLP tend to forget defectors.

\subsection{Trade-off in controlling variables between environments}

In order to keep the experiments between homogeneous and 
heterogeneous environments as controlled as possible, 
one has to endure a trade-off between using the same number of agents $N$
in total versus the same number of agents per $\rho$ and $s$,
that is 6 for homogeneous versus 1 for heterogeneous.

With the question of "what would happen if there were only one agent per
$\rho$ instead of 6 in homogeneous environment?" in mind,
we conducted the same experiment of homogeneous environment
with $N = 21$ instead.
We found that the result did not change compared to \reffig{fig:homogeneous}.
Therefore we simply elaborate on this experiment result instead of 
reporting a duplicate figure.

\subsection{Forgiveness is an Adaptation}
\label{subsec:ForgivenessIsAnAdaptation}

In a competitive, heterogeneous environment,
it is observed that forgetting defectors is clearly the superior forgetting strategy.
The reason for this is, if an opponent defects in the first encounter,
it is marked as a defector and is never played against until it is forgotten.
Since an agent that was defected only knows the defector
by means of perceived cooperation ratio, regardless of opponent's
actual cooperation probability, this is the case even if $\rho = 0.95$ for the opponent.
Consequently, the agent loses a potential cooperator who would cooperate most of the time
and would benefit the agent in the long term.
This is observed in \reffig{fig:heterogeneous} with FMD being consistently superior,
and in \reffig{fig:heatmap} 
where payoff ratio of cooperators of strategies FMU, FR, FLP, FMD with $\mu = 1$ 
being lower than those of $\mu = 0.95$ and $\mu = 0.90$
Therefore, in case of significant memory,
the protection that comes from keeping the defectors in memory
is not as significant as the potential gains that could be made by giving
the so called defectors second chances.

\section{Conclusion}

In this work, we go over the existing strategies in 
iterated prisoner's dilemma with memory game.
With the intention of representing real life better in a competitive way,
we extend these strategies by improving the existing ones,
introducing novel ones and running simulations on heterogeneous environment.
It is observed that in a more realistic environment consisting of all types of agents,
in terms of both cooperation probabilities and forgetting strategies,
agents who forget defectors consistently outperform other forgetting strategies 
for all memory ratio values.
Moreover, the best performing defectors are also the ones that forget other defectors.
In other words, agents who ``forgive'' defectors are the best performers.
Hence, forgiveness is an adaptation.

\section*{Acknowledgment}

We would like to thank Orhun Gorkem for constructive comments.
This work is partially supported by 
the Turkish Directorate of Strategy and Budget
under the TAM Project number 2007K12-873.

The code to create environments and run simulations is available at
\url{https://github.com/meliksahturker/Iterated-Prisoner-s-Dilemma-Framework}.

\bibliographystyle{ieeetr}
\bibliography{pFSinIPD-v01}
\end{document}